\documentclass[preprint,showpacs,preprintnumbers,amsmath,amssymb, showkeys,superscriptaddress,titlepage]{revtex4-2}
\usepackage{graphicx}
\usepackage{dcolumn}
\usepackage{bm}
\usepackage{hyperref}
\usepackage{textcomp}

\begin{document}

\title{COSMOPHYSICAL ASPECTS OF RELATIVISTIC NUCLEAR FRAGMENTATION}

\author{A.A. Zaitsev}
\email{zaicev@jinr.ru}
\affiliation{Joint Institute for Nuclear Research (JINR), Dubna, Russia}
\affiliation{P.N. Lebedev Physical Institute of the Russian Academy of Sciences (LPI), Moscow, Russia}
\author{N. Marimuthu}
\affiliation{Joint Institute for Nuclear Research (JINR), Dubna, Russia}
\author{D. A. Artemenkov}
\affiliation{Joint Institute for Nuclear Research (JINR), Dubna, Russia}
\author{P.I. Zarubin}
\affiliation{Joint Institute for Nuclear Research (JINR), Dubna, Russia}
\affiliation{P.N. Lebedev Physical Institute of the Russian Academy of Sciences (LPI), Moscow, Russia}
\author{N.G. Peresadko}
\affiliation{P.N. Lebedev Physical Institute of the Russian Academy of Sciences (LPI), Moscow, Russia}
\author{V. V. Rusakova}
\affiliation{Joint Institute for Nuclear Research (JINR), Dubna, Russia}

\begin{abstract}
The status of the study of multiple fragmentation of 950 MeV per nucleon Kr nuclei in a nuclear track emulsion aimed at determining the contributions of 2$\alpha$ decays of $^{8}$Be, the Hoyle 3$\alpha$ state, and the search for a 4$\alpha$ particle condensate state, is presented. In events with the production of few relativistic fragments of He and H, the possibility of estimating the multiplicity of neutrons in the fragmentation cone of a projectile nucleus is studied. For the planar component of neutron transverse momenta estimated from the angles of observed secondary stars, the Rayleigh distribution parameter was 35 $\pm$ 7 MeV/$c$. The importance of such events for the interpretation of cosmophysical observations is noted.
\end{abstract}

\pacs{21.60.Gx, 25.75.-q, 29.40.Rg}
\keywords{nuclear track emulsion, dissociation, invariant mass, relativistic fragments, $^{8}$Be nucleus, alpha particles} 
\maketitle 

\section{Introduction}
Starting with the discovery of the nuclear component of cosmic rays, only the nuclear track emulsion method (NTE) makes it possible to study the composition of the relativistic fragmentation of nuclei already at high-energy accelerators. The promising potential of the relativistic approach to the analysis of ensembles of fragments was manifested in NTE exposures to several GeV per nucleon nuclei which began at the JINR Synchrophasotron and Bevalac at the Lawrence Berkeley Laboratory (USA) as early as the 1970s.  However, objective difficulties prevented progress in this direction by other methods. Since the 2000s of the NTE method is applied in the BECQUEREL experiment at the JINR Nuclotron in respect to the cluster structure of nuclei, including radioactive ones, as well as the search for unstable nuclear-molecular states \cite{1}.

According to observations, in a nuclear emulsion, relativistic nuclei, up to the heaviest ones, can fragment into nucleons and nucleon clusters. The latter are represented by H and He isotopes (mainly protons and $\alpha$-particles), as well as light nuclei with a pronounced cluster structure. One of the most striking examples is shown in Fig. 1, where the breakdown of ionization as a result of multiple fragmentation of the projectile nucleus is clearly visible. A few percent of interactions contain jets of fragments with a total charge close to the charge of the initial nucleus, often not accompanied by the formation of target fragments. So far, no interpretation of this phenomenon has been proposed. In addition, secondary stars are often observed inside the fragmentation cone of heavy nuclei, which do not have tracks of interaction vertices that form relativistic fragment neutrons.

The average range of relativistic nucleons in the NTE medium is about 35 cm. In principle, the neutron multiplicity can be estimated from the multiply decreasing average length of the appearance of nuclear stars that do not have an entrance track. Being laborious, measurements of multiple trace scattering make it possible to identify the relativistic isotopes $^{2,3}$H and $^{3,4}$He, which indirectly indicate the number of neutrons in the most peripheral collisions. A holistic study of neutron production in an antilaboratory system is important both in nuclear astrophysics and space physics and in applied nuclear research.

The BECQUEREL experiment aims to search in the relativistic approach for the $\alpha$-particle Bose-Einstein condensate ($\alpha$BEC), an unstable state of $S$-wave $\alpha$-particles \cite{2,3}. The extremely short-lived $^{8}$Be nucleus is described as 2$\alpha$BEC, and the $^{12}$C(0$^+_2$) excitation or Hoyle state (HS) as 3$\alpha$BEC. The realizability of more complex states is essential in nuclear astrophysics. The search for $\alpha$BEC is carried out in NTE layers longitudinally exposed to nuclei from C to Au. The invariant mass of $\alpha$-ensembles is determined from the emission angles in the approximation of conservation of momentum per nucleon of the parent nucleus. For light nuclei, it was shown that, due to the extremely small energies and widths, the decays of $^{8}$Be and HS are identified by its limitation. The $^{8}$Be $\to$ 2$\alpha$ and $^{12}$C(0$^+_2$) $\to$ $^{8}$Be$\alpha$ decays can serve as signatures for more complex $n\alpha$BEC decays. Thus, the $^{16}$O(0$^+_6$) state at 660 keV above the 4$\alpha$ threshold, considered as 4$\alpha$BEC, can decay into $\alpha^{12}$C(0$^+_2$) or 2$^{8}$Be.

Having been tested with light nuclei, the universal invariant mass method has been used to identify $^{8}$Be and HS and to search for more complex $n\alpha$BEC states within the previously accumulated data on the fragmentation of medium and heavy nuclei. Recently, based on the statistics of dozens of $^{8}$Be decays, an increase in the probability of detecting $^{8}$Be with an increase in the number of associated $\alpha$-particles $n\alpha$ was found. A preliminary conclusion is drawn that the contributions from $^{9}$B and HS decays also increase. The exotically large sizes and lifetimes of $^{8}$Be and HS suggest the possibility of synthesizing $\alpha$BEC by sequentially picking up emerging $\alpha$-particles 2$\alpha$ $\to$ $^{8}$Be, $^{8}$Be$\alpha$ $\to$ $^{12}$C(0$^+_2$), $^{12}$C(0$^+_2$)$\alpha$ $\to$ $^{16}$O(0$^+_6$), 2$^{8}$Be $\to$ $^{16}$O(0$^+_6$) and further with a decreasing probability at each step, when $\gamma$-quanta or recoil particles are emitted. The possibility of indirect determination of the cross section of these fusion reactions from the probabilities of dissociation into unstable states deserves to be studied.

The following is the status of the $\alpha$BEC search and the first observations of neutron production in the BECQUEREL experiment.

\begin{figure}
	\centerline{\includegraphics*[width=1.0\linewidth]{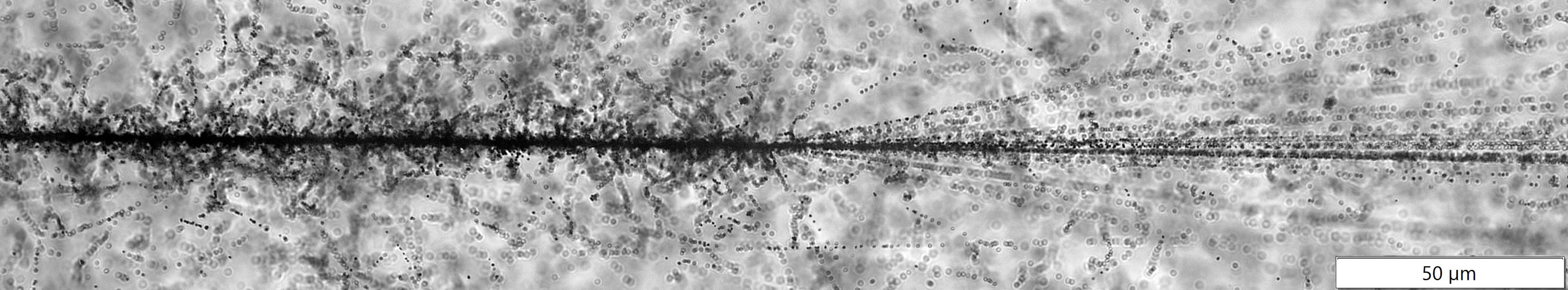}}
	\caption{Macrophotograph of peripheral interaction of 3.8 GeV per nucleon $^{124}$Xe nucleus (exposed in 2022 at NICA/Nuclotron).}
\end{figure}

\begin{figure}
	\centerline{\includegraphics*[width=.65\linewidth]{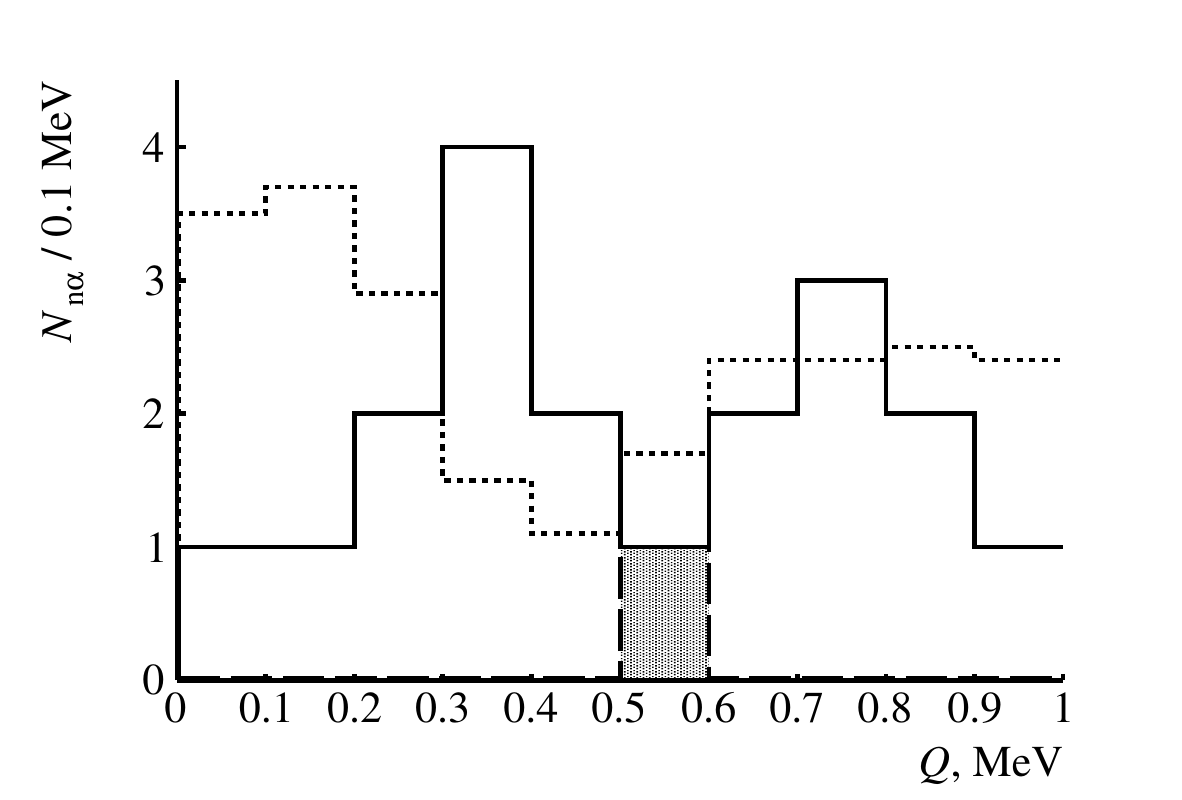}}
	\caption{Distribution in range of small values of invariant mass $Q$ of pairs (points), triplets (solid line) and quadruples (shaded line) of $\alpha$-particles formed in the fragmentation of Kr nuclei.}
\end{figure}

\section{Relativistic $\alpha$-particle states}

In the BECQUEREL experiment, a purposeful study of $n\alpha$-channels of $^{84}$Kr fragmentation of up to 1 GeV per nucleon in nuclear emission layers irradiated longitudinally in the 1990s is being carried out in GSI (Darmstadt) \cite{1}. 
An accelerated search for such interactions with a transverse view of the layers was started on a motorized microscope Olympus BX63. The statistics of measured events with the formation of 5-10 $\alpha$-particles in the fragmentation cone reached 146 in 6 layers.
On Fig. 2 shows the distributions over the invariant mass $Q$ of pairs, triplets, and quadruples of $\alpha$-particles. 
The number of events with $^{8}$Be ($Q_{2\alpha}$ $\textless$ 0.4 MeV) decays is 96, HS ($Q_{3\alpha}$ $\textless$ 0.7 MeV) - 13, and 2$^{8}$Be – 17, giving the ratio HS/$^{8}$Be 0.14 $\pm$ 0.04, and 2$^{8}$Be - 0.18 $\pm$ 0.05. 
Their distribution over $n\alpha$ \cite{1} demonstrates mutual proportionality, therefore, the growth when normalized to the number of events $n\alpha$ rapidly decreases with $n\alpha$. The 2$^{8}$Be event ($Q_{4\alpha}$ = 0.6 MeV) isolated in the region $Q_{4\alpha}$ $\textless$ 1 MeV was identified.

The collected statistics make it possible to reveal the neutron component, which increases with $n\alpha$. At the same time, the multiplicity of spectator neutrons emitted in interactions should increase even more rapidly with the multiplicity of associated H fragments (mainly protons). On Fig. 3 shows the correlation of He and H, where the average values are 5.3 $\pm$ 0.1 and 4.3 $\pm$ 0.2, respectively. The total charge of these fragments is about 18, i.e., approximately half of the charge of the parent nucleus. Thus, one can expect instantaneous emission of several tens of neutrons remaining in the fragmentation cone.

\begin{figure}
	\centerline{\includegraphics*[width=.65\linewidth]{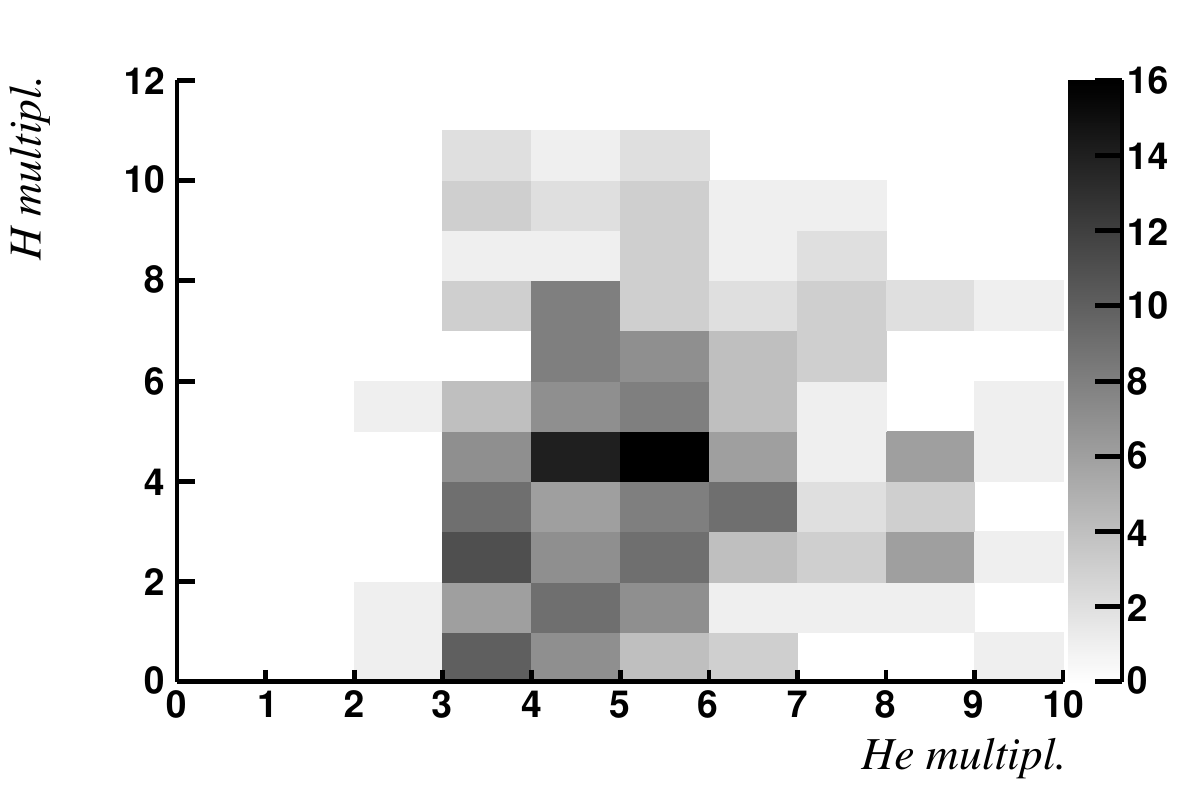}}
	\caption{Distribution of events over multiplicity of relativistic He and H fragments.}
\end{figure}

\begin{figure}
	\centerline{\includegraphics*[width=.75\linewidth]{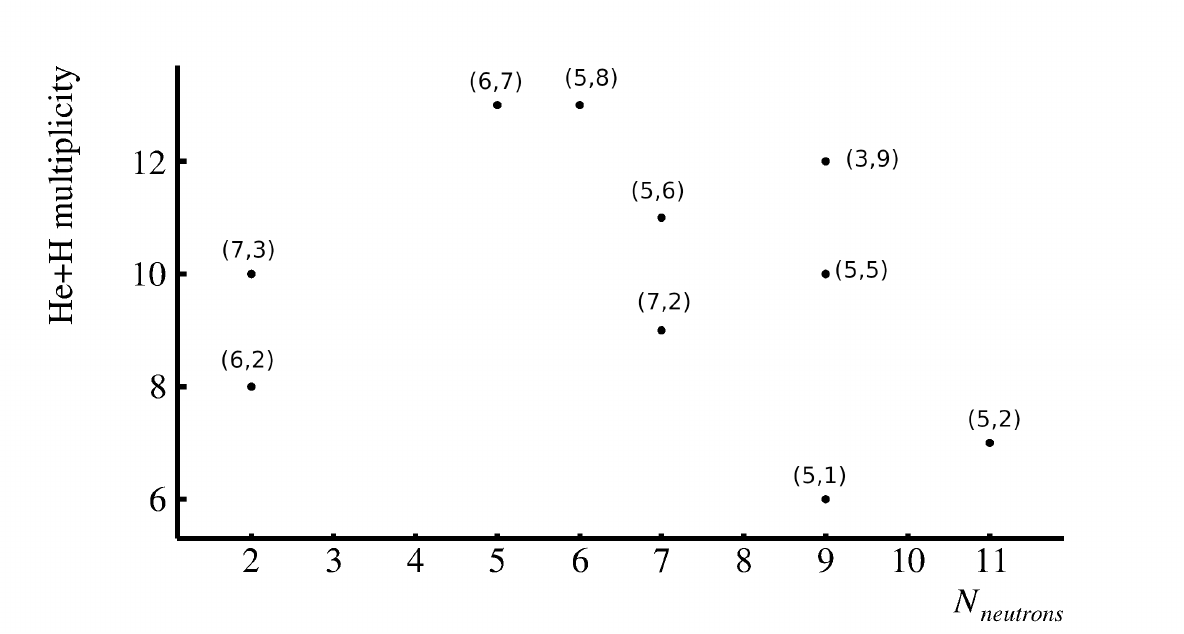}}
	\caption{Correlation of SNS number on multiplicity of He and H fragments in fragmentation cone; numbers of He and H fragments are given in parentheses.}
\end{figure}

\section{Observation of secondary neutrons}

To gain experience in searching for secondary stars in the fragmentation cone, which could be neutron-induced (SNS), 10 events were used, from among those described above, with the topology (He, H) marked in Fig. 4. In the selected 10 events, the average range of Kr nuclei in the nuclear emission before inelastic interaction was 19 mm. When viewed approximately 10 mm from the vertices in X and $\pm$ 0.3 mm in Y, 67 SNS were found, examples of which are shown in Fig. 5. Distribution of SNS in Fig. 4 can serve to develop a probabilistic estimate of the neutron multiplicity.

According to the distribution in the plane of the NTE layer along the longitudinal X and transverse Y coordinates from the vertices (Fig. 6), SNS are concentrated on the area up to 4 mm along X and $\pm$ 0.2 mm along Y. A further decrease in their number can be determined by the reduction of the enclosed solid angle. Without significantly reducing the SNS statistics, the area constraint makes it possible to reduce the possible background. Measurements of the Z coordinates by depth of field are started.

On Fig. 7 shows the planar angle distribution between the direction of the primary track Kr and the registered event SNS in the focal plane XY. The resulting distribution is described by a Gaussian function with parameters $\left\langle \phi \right\rangle $  = -18 $\pm$ 1 mrad and $\sigma$ = 38 $\pm$ 1 mrad. In the approximation of conservation of momentum per nucleon Fig. 8 shows the distribution of the planar component of the transverse momentum $p_y$ of neutrons. It is described by the Rayleigh function with the parameter $\sigma$ = 35 $\pm$ 7 MeV/$c$. Then the average transverse momentum, taking into account the volume factor, is 50 MeV/$c$, corresponding to energy in the system of the parent nucleus of 1.3 MeV. Such a small value indicates the decisive contribution of the momentum distribution of peripheral neutrons in the parent nucleus. It can be used for estimations in cosmophysical and applied research.

\begin{figure}
	\centerline{\includegraphics*[width=.75\linewidth]{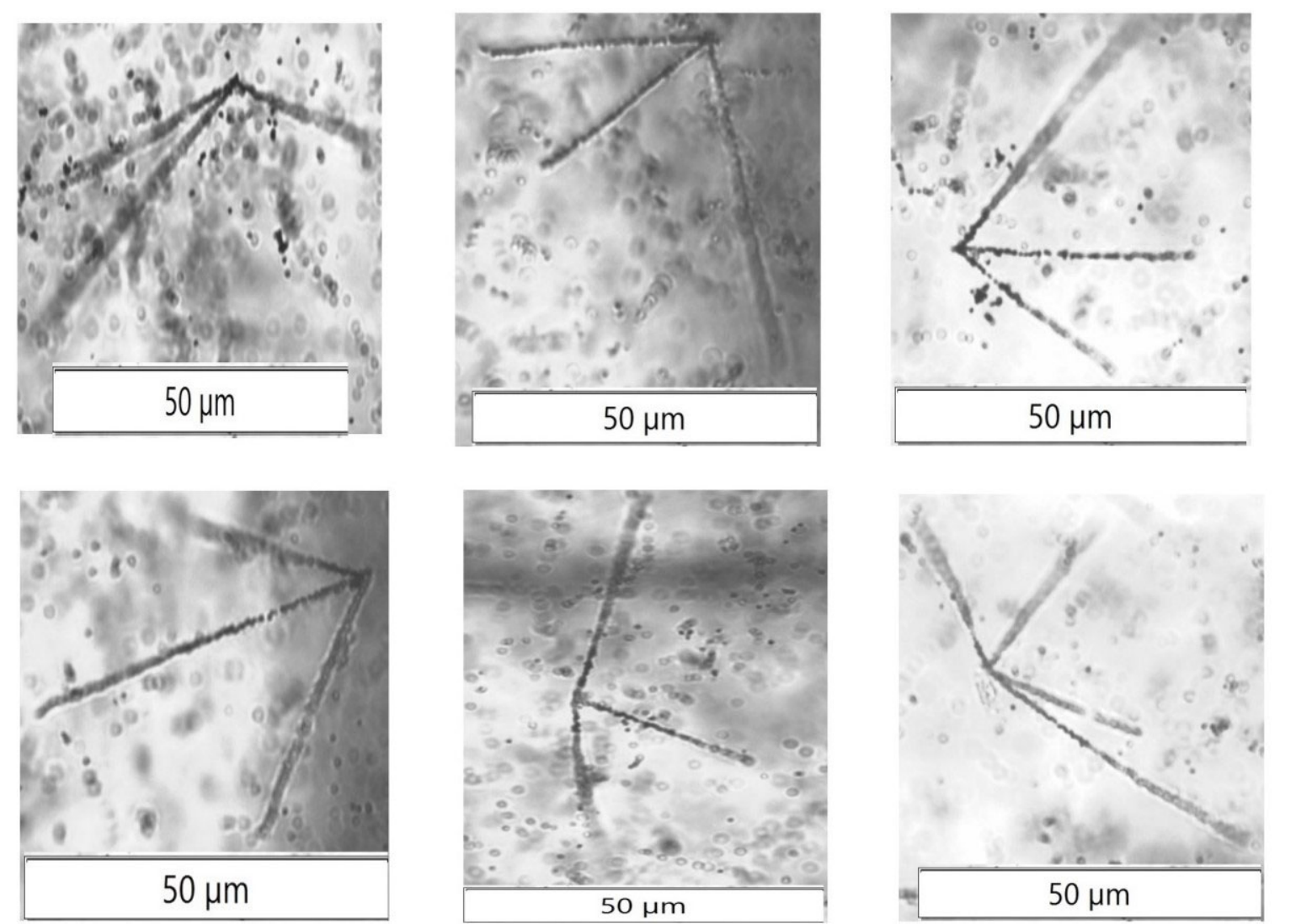}}
	\caption{Macrophotographs of secondary stars in fragmentation cone, interpreted as induced by neutrons (SNS).}
\end{figure}

\begin{figure}
	\centerline{\includegraphics*[width=.75\linewidth]{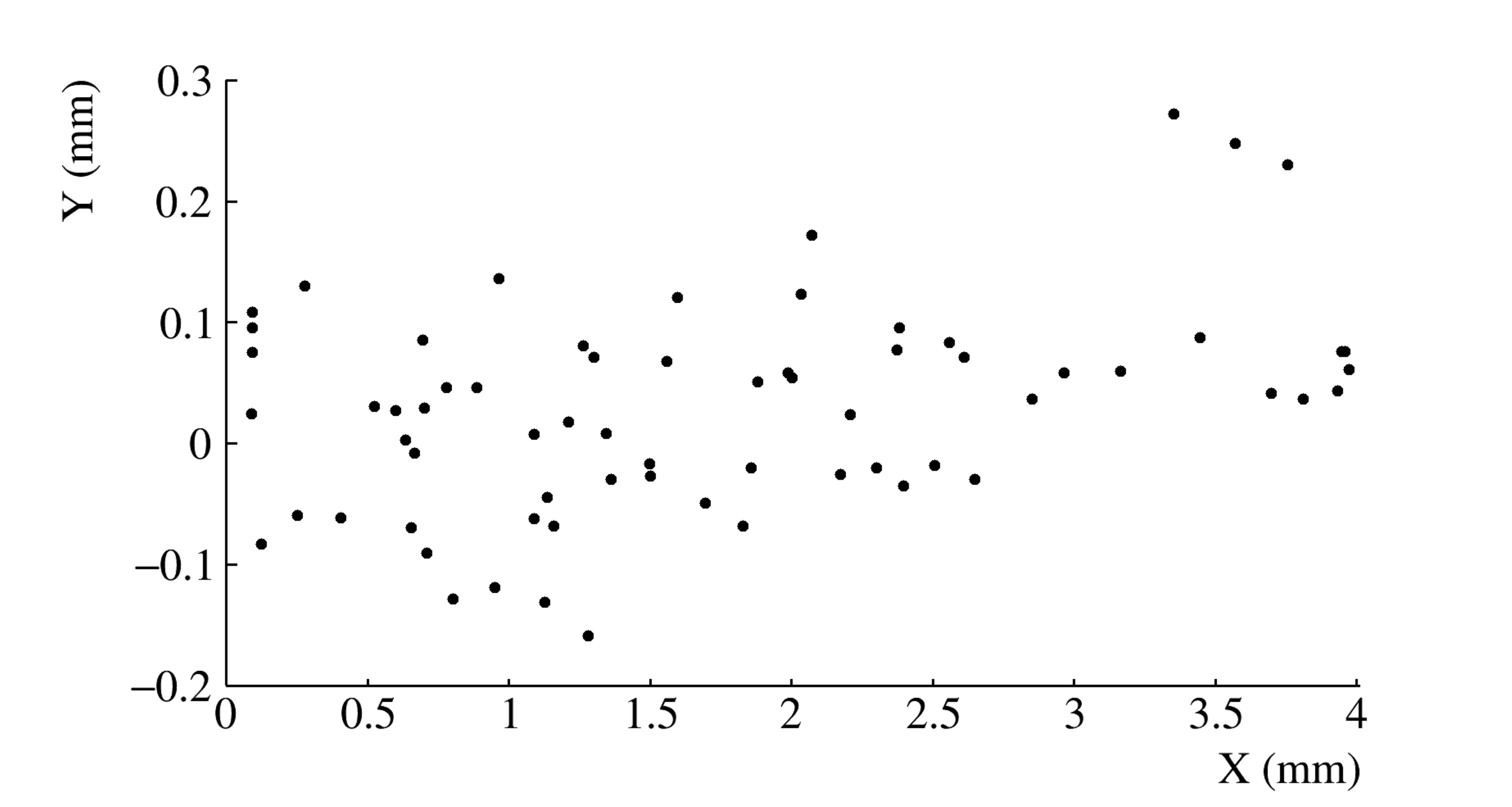}}
	\caption{Distribution of SNS coordinates in XY focal plane relative to primary vertex.}
\end{figure}

\begin{figure}
	\centerline{\includegraphics*[width=.8\linewidth]{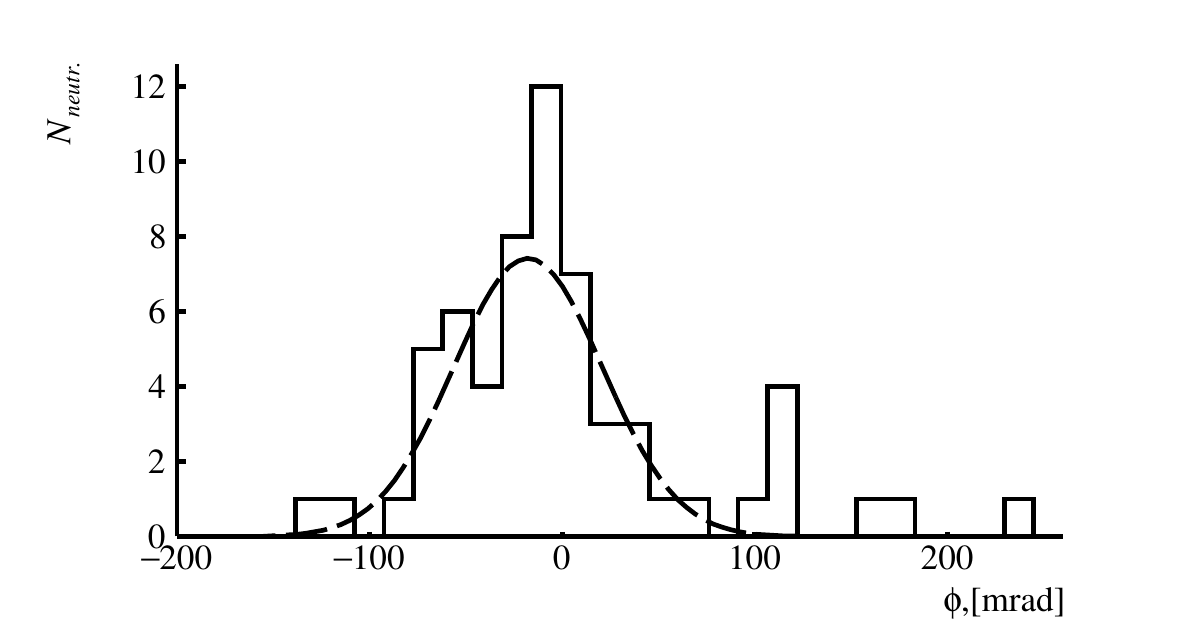}}
	\caption{Distribution over flat angle $\phi$ of observation of the SNS with respect to direction of primary track, approximated by Gaussian function (dotted line).}
\end{figure}

\section{CONCLUSION}
Being a source for the search for unexplored forms of nuclear matter, the fragmentation of heavy nuclei with the simultaneous formation of tens of nucleons and the lightest nuclei should also manifest itself in cosmophysical aspects. Interaction in the Earth's atmosphere of heavy nuclei of cosmic origin can lead to events that are effective sources of secondary mesons and, as a result, groups of relativistic muons (muon bundles). Their passage of the latter is noted in large-scale detectors at colliders (for example, \cite{4}). Observations of muon groups with large angles with respect to the zenith are carried out in the NEVOD-DECOR experiment (recent publications \cite{5,6,7}).

Giant energies of primary nuclei do not significantly change the course of dissociation near bond thresholds, which essentially remains a low-energy phenomenon. However, they lead to compression of the fragmentation cone by the corresponding orders of magnitude and collimation of detected muons. Another question is to what extent the multiple fragmentation events simulate the interaction of protons with energy one or two orders of magnitude greater than that of nuclear nucleons. To take this phenomenon into account, it will be necessary to supplement the pattern of successive fragmentation with the instantaneous production of the lightest nuclei and neutrons. The presented study can serve as a starting point in this direction.

\begin{figure}
	\centerline{\includegraphics*[width=.75\linewidth]{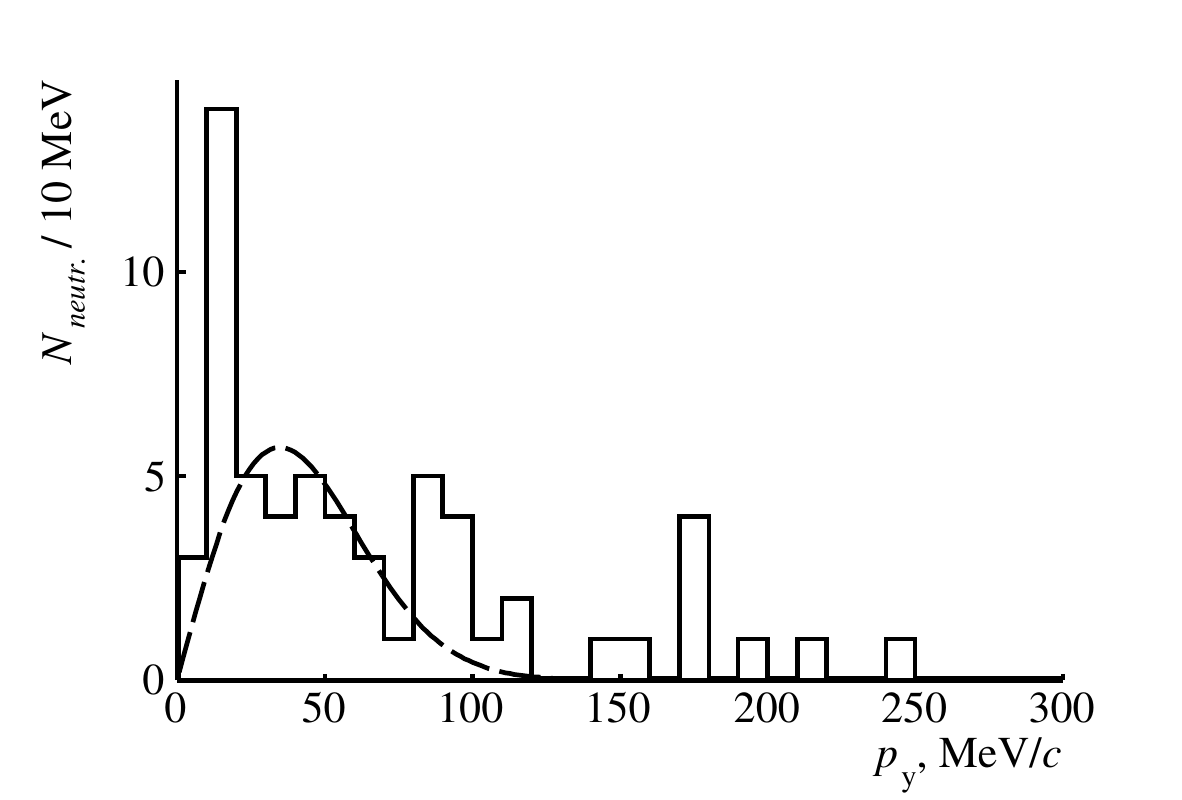}}
	\caption{Planar component distribution of neutron transverse momenta in the fragmentation cone, approximated by Rayleigh function (dashed).}
\end{figure}

The low energy of nuclei (significantly below the limiting fragmentation regime) leads to their rapid deceleration and complicates the dynamics of interactions, which impairs the identification of HS and, thus, its prospects for 4$\alpha$BEC \cite{3}. Well-established approaches can be applied by irradiating nuclear emission with heavy nuclei at an energy of several GeV per nucleon. Narrowing of the fragmentation cone also turns out to be useful in relation to the observation of neutron-induced stars. This possibility was opened up by the acceleration of $^{124}$Xe nuclei to an energy of 3.8 GeV per nucleon, carried out in the winter run of 2022 at the NICA/Nuclotron accelerator complex. The analysis of the stacks of NTEs irradiated in the initial section of the extracted beam and in the building of extracted beams 205 after the BM@N experiment has been started. In the latter case, a beam of heavy nuclei is brought along a magneto-optical channel about 70 meters long in an ion pipeline.

The search for $n\alpha$BEC serves as a starting point for studying nuclear matter with temperatures and densities from red giants to supernovae. In this regard, the nuclear emission layers exposed to heavy nuclei at several GeV per nucleon of the NICA accelerator complex will allow applying proven approaches to the analysis of relativistic ensembles of H and He isotopes, as well as neutrons, of unprecedented multiplicity. The ratios and values of the energy in the system of the projectile nucleus of the identified isotopes $^{1,2,3}$H and $^{3,4}$He characterize the emerging matter \cite{8}.

%
%


\begin{thebibliography}{REFERENCES}
%

\bibitem {1}
D. A. Artemenkov, V. Bradnova, O. N. Kashanskaya, N. V. Kondratieva, N. K. Kornegrutsa, E. Mitsova, N. G. Peresadko, V. V. Rusakova, R. Stanoeva, A. A. Zaitsev, I. G. Zarubina, P. I. Zarubin, Phys. At. Nucl. \textbf{85}, 528 (2022); \href{doi.org/10.1134/S1063778822060035}{doi.org/10.1134/S1063778822060035};
\href{https://arxiv.org/abs/2206.09690}{arXiv: 2206.09690}.

\bibitem {2}
A. Tohsaki, H. Horiuchi, P. Schuck, and G. Ropke, Rev. Mod. Phys. \textbf{89}, 011002 (2017); \href{doi.org/10.1103/RevModPhys.89.011002}{doi.org/10.1103/RevModPhys.89.011002}.

\bibitem {3}
W. von Oertzen, Lect. Notes Phys. \textbf{818}, 1 (2010); \href{doi.org/10.1007/978-3-642-13899-7_3}{doi.org/10.1007/978-3-642-13899-7\_3}.

\bibitem {4}
J. Abdallah, P. Abreu, W. Adam, P. Adzic, T. Albrecht, T. Alderweireld, R. Alemany-Fernandez, T. Allmendinger, P. P. Allport, U. Amaldi, N. Amapane, S. Amato, E. Anashkin, A. Andreazza, S. Andringa, N. Anjos et al.  et al. 
(DELPHI Collaboration), Astophys. J. \textbf{28}, 273 (2007); 
\href{doi.org/10.1016/j.astropartphys.2007.06.001}{doi.org/10.1016/j.astropartphys.2007.06.001};
\href{https://arxiv.org/abs/0706.2561}{arXive: 0706.2561 [astro-ph]}.

\bibitem {5}
A.G. Bogdanov, R.P. Kokoulin, G. Manocchi, A.A. Petrukhin, O. Saavedra, Astophys. J. \textbf{98}, 13 (2018); \href{doi.org/10.1016/j.astropartphys.2018.01.003}{doi.org/10.1016/j.astropartphys.2018.01.003}.

\bibitem {6}
V.S Vorobev, A.A. Petrukhin, Phys. At. Nucl. \textbf{84}, 934 (2021); \href{doi.org/10.1134/S1063778821130408}{doi.org/10.1134/S1063778821130408}.

\bibitem {7}
G. Trinchero, M.B. Amelchakov, A.G. Bogdanov, A. Chiavassa, A.N. Dmitrieva, G. Mannocchi, S.S. Khokhlov, R.P. Kokoulin, K.G. Kompaniets, A.A. Petrukhin, V.V. Shutenko, I.A. Shulzhenko, I.I. Yashin, E.A. Yurina, Astophys. J. \textbf{945}, 123 (2023); \href{doi.org/10.3847/1538-4357/acb1fc}{doi.org/10.3847/1538-4357/acb1fc}; 
\href{https://arxiv.org/abs/2210.09690}{arXiv: 2210.09690}.

\bibitem {8}
S. Typel, G. Röpke, T. Klähn, D. Blaschke, H. H. Wolter, Phys. Rev.  C \textbf{81}, 015803 (2010); \href{doi.org/10.1103/PhysRevC.81.015803}{doi.org/10.1103/PhysRevC.81.015803}; 
\href{https://arXiv.org/abs/0908.2344}{arXiv 0908.2344 [nucl-th]}.

\end{thebibliography}
\end{document}